\documentclass[10pt]{article}
\usepackage[dvips]{graphics, graphicx}
\usepackage{latexsym,epsf}
\usepackage{multicol}
\usepackage[centertags]{amsmath}

\fontencoding{OMS}\selectfont

\def\thebibliography#1{\setlength{\itemsep}{0pt}\setlength{\parsep}{0pt}
\refname{\small{\textbf{}}}
  \list
  {\arabic{enumi}.}{\settowidth\labelwidth{[#1]}\leftmargin\labelwidth
   \advance\leftmargin\labelsep
   \usecounter{enumi}}
\let\endthebibliography=\endlist}
\renewcommand\refname{\textbf{{REFERENCES}}}
\title{\textbf{COULOMB SUMS FOR  $^7$Li NUCLEUS
AT  3-MOMENTUM TRANSFERS  $q = 1.250 - 1.625 \:fm^{-1}$ }}
\author{\textbf{\textit{A.Yu. Buki\footnote{\normalfont
Corresponding author. E-mail address: abuki@ukr.net.ua}, N.G.
Shevchenko, I.S. Timchenko }}
\\
\emph{\small National Scientific Center "Kharkov Institute of
Physics and Technology", 61108, Kharkov, Ukraine}\\
{\small(Received June 1, 2008)}}
\begin{document}
\date{}
\maketitle
\begin{center}
\begin{minipage}{165 mm}
{\small
The experimental response functions of  $^7$Li nucleus at effective
3-momentum transfers  $q = 1.250;\: 1.375;\: 1.500$ and $1.625
\:fm^{-1}$ are presented. The longitudinal response functions were
used to evaluate the Coulomb sum values. The Coulomb sums for $^6$Li
 obtained by us earlier were applied to analyze these data. The
Coulomb sums of lithium isotopes were compared with the well-known
Coulomb sums values of the other nuclei.
}
\end{minipage}
\end{center}
\begin{multicols}{2}
\begin{center}
\textbf{\textsc{1. INTRODUCTION}}
\end{center}
The longitudinal ($R_L$) and transverse ($R_T$) response functions
represent the spectra of scattered electrons separated into
longitudinal and transverse components respectively according to
polarization of electromagnetic-interaction field.
 The relation between the response functions (RF) and the doubly
differential electron-scattering cross section
($\rm{d^2}\sigma/\rm{d}\Omega\rm{d}\omega$), according to
ref.~\cite{de Forest}, can be written as
\begin{equation}
\frac{\rm{d}^2\sigma}{\rm{d}\Omega\rm{d}\omega}\left(\theta,E_0,\omega\right)/\left(\sigma_M
(\theta,E_0)\right) \nonumber\\
\end{equation}
\begin{equation}
=\frac{q^4_\mu}{q^4} \cdot R_L(q,\omega)+ \left[\frac{1}{2} \cdot
\frac{q^2_\mu}{q^2}+\tan^2\frac{\theta}{2}\right] \cdot
R_T(q,\omega), \label{diseqn1}
\end{equation}
where $E_0$ is the initial energy of electron scattered through the
angle $\theta$ with the transfer of energy  $\omega$, effective
3-momentum \mbox{$q = \xi \cdot \{ 4E_0[E_0 - \omega]$
$\sin^2(\theta/2) + \omega^2 \}^{1/2}$} and 4-momentum \mbox{$q_\mu
= (q^2-\omega^2)^{1/2}$} to the nucleus involved;
$\sigma_M(\theta,E_0) = Z^2 e^4 \cos^2 ( \theta/2)/ [4E_0^2
\sin^4(\theta /2) ]$ is the Mott cross section, $e$ is the electron
charge. The correction $\xi$ takes into account the distortion of
the electron wave by the electrostatic field of nucleus. According
to \cite{Yennie}, this correction is written as \mbox{$\xi$ = 1 +
1.33$Ze^2/(E_0<r^2>^{1/2})$}, where $Z$ and \mbox{$<r^2>$} are,
respectively, the charge and r.m.s. radius of the nucleus.

At the present time  the theoretical calculations of
$R_{T/L}$-functions are rather difficult and exist only for nuclei
with $A \leq 4$. Therefore, the experimental data are presented as
RF moments, which are compered with calculation by the sum-rule
approach.  The moment of RF have the following form
\begin{equation}
S^{(n)}_{T/L}(q)=\frac{1}{Z}\int_{\omega^+_{el}}^{\infty}\frac{R_{T/L}(q,\omega)}{\eta
\cdot G^2(q^2_\mu)} \cdot \omega^n \rm{d}\omega, \label{diseqn2}
\end{equation}
where $n$ is the moment number, $G(q^2_\mu)$ is the electric form
factor of the proton; \mbox{$\eta = [1+q^2_\mu /
(4M^2)]\times[1+q^2_\mu / (2M^2)]^{-1}$} is the correction for the
relativistic effect of nucleon motion in the nucleus; $M$ is the
proton mass; $\omega_{el}^+$ means that the bottom boundary of the
integration domain is the energy transferred that corresponds to
elastic scattering of the electron from the nucleus. But the
integral does not include the elastic scattering form factor.

Usually  the $R_L$-function moment with $n=0$ is obtained from the
measurements of RF. It is named Coulomb Sum (CS) and denoted as
$S_L(q)$.

The investigation of the CS isotopic differences of $^6$Li and
$^7$Li nuclei was the original aim of our measurements. However, as
the result of the processing of only part of the experimental data,
the interesting features of $^7$Li CS values were discovered. The
present paper deals with these CS features.

 \vspace{3mm}
\begin{center}
\textbf{\textsc{2. EXPERIMENTAL DATA}}
\end{center}

The spectra of electrons scattered by $^7$Li nuclei were obtained at
the linear accelerator LUE-300 of NSC KIPT at initial energy $E_0 =
129$ to $259\:MeV $ and scattering angles $\theta = 60^\circ 30'$
 to $94^\circ 10'$, $\theta = 160^\circ$. The range of the
measurements of the 3-momenta and energies transferred to nuclues
are shown in fig.1.

The experimental equipment and the measurement method have been
described in refs.~\cite{Buki_2002,Buki_1995}. The data processing
and the error analysis were performed as in
refs.~\cite{Buki_1995,Buki_2006}. In regard to the last we note that
this question has been given some consideration in the paper,
because the errors of the experimental RF and, consequently, the
errors of CS significantly depend on the systematical errors of the
absolutization of the measured cross sections. Then, before and
after the measuring of each spectrum of electrons scattered by
$^7$Li, the measurements of the $^{12}$C ground state form factor
were carried out. The absolutization of the measured
$^7$Li$(e,\:e')$ cross sections was performed through the comparison
of these data and the particularly precise values of $^{12}$C form
factor from ref.~\cite{Reuter}. At the same time the correction
obtained in ref.~\cite{Buki_2007} was applied to data of
ref.~\cite{Reuter}. As additional verification the comparison of the
measured during the experiment $^7$Li ground state form factor and
its magnitude from ref.~\cite{Suelzle} was done.

\begin{center}
\includegraphics[width=0.50\textwidth]{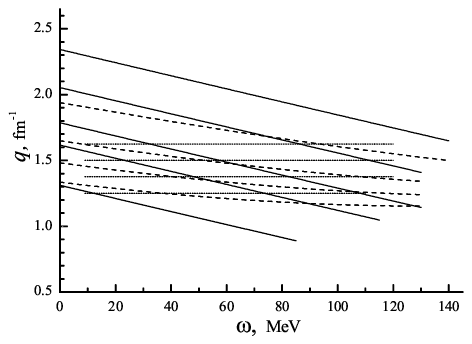}
 \textbf{\it Fig.1.} {\it The transferred 3-momenta and energies
 of the electron scattered spectra. The solid lines label the measured
 at $\theta = 160^\circ$ spectra, the dashed lines show the measurements at $\theta =
60^\circ 30'$ to $94^\circ 10'$, the dotted lines are the constant
values of the transferred  3-momenta, at which the RF are obtained
 }\\
\end{center}

As a result of the data processing through the usage of eq.~1, the
$R_{T/L}$-function values for $^7$Li nucleus at $q = 1.250 - 1.625\:
fm^{-1}$ were obtained. For instance, RF at $q=1.375 fm^{-1}$ is
present in fig.2. It is evident from fig.~2  that to determine
experimentally CS, it is essential that RF should be integrated to $
\omega = \infty$. For this purpose RF were extrapolated  by the
 function $R \propto \omega ^{-\alpha}$ (see refs.~\cite{Tornow,
 Orlandini}) to the region where the measurements are impossible. The value $\alpha = 2.45 \pm
 0.15$ of the $^7$Li longitudinal RF was
 found by the method of ref.~\cite{Buki_2005}.
 The obtained in such a way CS values are shown in fig.~3.
 The shown in the figure errors are statistical.

 First of all the characteristic feature of these data is that  the average
 value of  $^7$Li $S_L(q)$ is equal to  $1.018 \pm 0.025 \pm 0.029$
 (the first error is statistical,
 and the second is systematical) at the transferred 3-momenta region
   $q = 1.375 - 1.625\: fm^{-1}$,
 while for nuclei with $Z > 1$ $S_L(q)$ it is less than  $0.8$ at $q = 1.5 \:fm^{-1}$
 (see, for instance, ref.~\cite{Zghiche}). To consider this phenomenon it is necessary
 to make sure of its validity. In this connection we
 note the following:
\begin{itemize}
    \item {At the same time, when the electron scattered by $^7$Li spectra were
    measured,
    we carried out the measurements of $^4$He$(e,e')$ spectra. The obtained from these data CS
    of $^4$He were in good agrement both with experimental Bates and Saclay data,
    and with theoretical calculations (see ref.~\cite{Buki_2006}).
    Consequently it seems to be improbable that the gross error is present
    in $^7$Li data.}
    \item {Simultaneous with $^7$Li the measurements of $^6$Li$(e,e')$ spectra were
    carried out. From general considerations the CS of lithium isotopes
    should not differ significantly. In spite of the fact, that not all $^6$Li data have been processed, some estimations of $^6$Li CS  may be done.
     At $q = 1.25\: fm^{-1}$ this estimation
     showed that the CS values of the lithium isotopes are close (see fig.~4).}
\end{itemize}
\end{multicols}

\begin{center}
\includegraphics[width=0.95\textwidth]{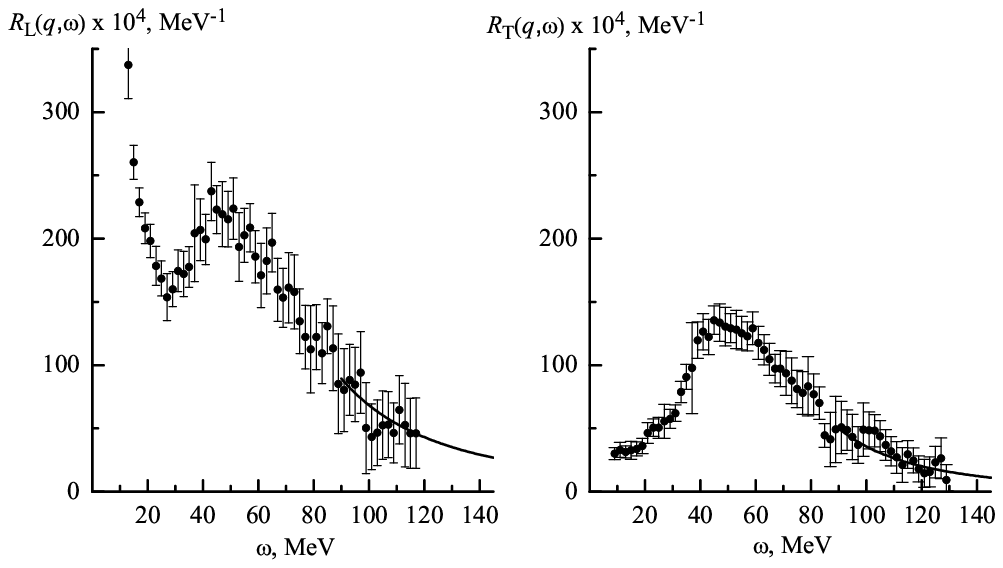}
 \textbf{\it Fig.2.} {\it {The longitudinal and transverse $^7$Li response functions
 at $q = 1.375\: fm^{-1}$.

 The solid lines show the extrapolations of
 RF (see the text)}
 }\\
\end{center}

\begin{multicols}{2}
\begin{itemize}
    \item {Before the measurements with $^6$Li and $^7$Li we had carried out the first measurements
    with $^6$Li \cite{Buki_1977}. The $^6$Li CS from ref.~\cite{Buki_1977} are denoted as
    $\sigma_l(q)$ and in the term of $\sigma_l(q)$ the modern determination of CS
    can be written as $S_L(q) = \sigma_l(q)/G^2(q^2)$.  $^6$Li CS values from ref.~\cite{Buki_1977} transformed in the same way are shown
    in fig.4\footnote{\normalfont
 It is necessary to say, that the characteristic features of $^7$Li CS values discussed here  may
be observed in the $^6$Li CS values also. However, in 1977, when
ref.~\cite{Buki_1977} was published, the obtained $^6$Li CS values
were nothing to compare with. At that time the systematical data of
CS values for the various nuclei were absent. The systematics
appeared as a result of Bates and Saclay works only after 1979.}.}
\end{itemize}

It is evident from fig.~4 that all available data for lithium
isotopes CS data are agree with each other. It is the basis to
consider the reliability of the obtained $^7$Li CS values as
sufficiently authorized.

\begin{center}
\includegraphics[width=0.50\textwidth]{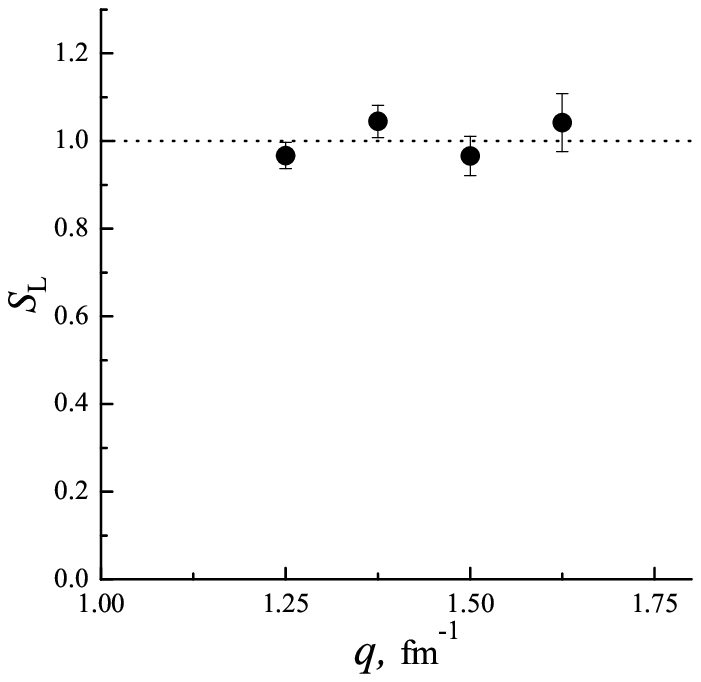}
 \textbf{\it Fig.3.} {\it The $^7$Li Coulomb Sums obtained in the present paper
 }\\
\end{center}

 \vspace{5 mm}
\begin{center}
\textbf{\textsc{3. DISCUSSION AND CONCLUSIONS}}
\end{center}

The $S_L(q)$ dependence shapes for $A > 2$ nuclei (with the
exception of the lithium isotopes) are similar with each other: at
transferred 3-momentum region $q = 0 - 2 \: fm^{-1}$ the smooth rise
with the increasing $q$ is observed, and at $q \geq 2\: fm^{-1}$
$S_L(q)$ it is equal to constant value (plateau is obtained). Let us
denote $S_L(q)$ in the plateau region as $S_{L,max}$. The value
$S_{L,max}$ is equal to 1.0 for $A \leq 3$ nuclei \cite{Dytman,Dow}.
In the case of all investigated in Bates and Saclay nuclei $A \geq
4$ the $S_{L,max}$ values decrease with the increase of atomic
number: from $0.9 \pm 0.03$ for $^4$He to $0.5 \div 0.6$ for
$^{208}$Pb (the effect of the Sum rule
quenching)\footnote{\normalfont
 Notice, that the attempt to solve the problem of the
 Sum rule quenching via introduction
 the corrections into the experimental data was made in
 ref.~\cite{Jourdan}. Thus in this work the $S_{L,max}$ values of $^{12}$C, $^{40}$Ca
 and $^{56}$Fe nuclei were observed to be closed 1, we think that
 work~\cite{Jourdan} is mistaken. The same conclusion was made by
 authors of ref.~\cite{Morgenstern}.}. As an illustration the straight
 line approximation of the experimental CS values of $^4$He is showed in fig.~4.

As it is seen from fig.4 the $S_{L}(q)$ dependencies of lithium
isotopes and $^4$He ones differ from each other and, as was
mentioned, from other nuclei. Let us discuss the following features
of lithium nuclei CS value.

\begin{center}
\includegraphics[width=0.50\textwidth]{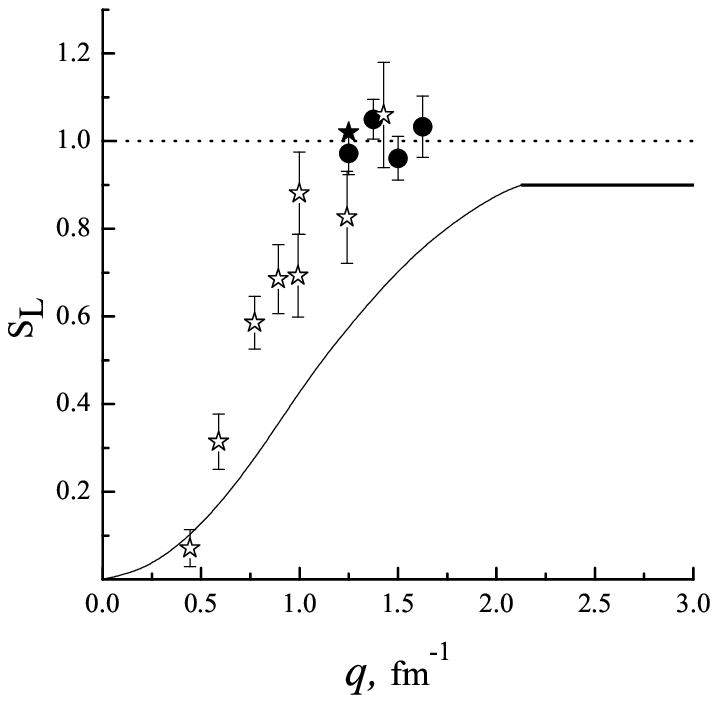}
 \textbf{\it Fig.4.} {\it The comparison of Coulomb sums of $^7$Li, $^6$Li
 and $^4$He nuclei. The CS values of $^7$Li are labeled as full circles,
 open stars show $^6$Li CS from ref.~\cite{Buki_1977},
 full star shows the $^6$Li CS value which is obtained from the data
 measured simultaneously with $^7$Li  data. The solid line shows the $^4$He
 data approximation: at $q < 2\: fm^{-1}$ the calculations of works
 \cite{Shiavilla} and \cite{Efros} are in good agrement with each other;
 at $q \geq 2\: fm^{-1}$ straight line shows the approximation of the CS values obtained
 in Bates and Saclay labs \cite{Buki_1977}
  }\\
\end{center}
\vspace{5 mm}

\begin{itemize}
    \item {The $S_{L}(q)$ dependence is equal to constant value already
    at $q = 1.25\: fm^{-1}$,  but in the case of other nuclei the it
    is equal to constant value only at $q \approx 2\: fm^{-1}$. This
    phenomenon is probably explained by the fact that lithium isotopes
    are very cauterized, while there are investigations of noclustered
    nuclei only in the systematic of $S_{L}(q)$.\footnote{\normalfont
    Using the results of the measurements of the $^6$Li $S_L(q)$ the clusterization parameter
    of this nucleus was obtained in ref.~\cite{Buki_1977} and its value was agreed
    with the result using the $(e, e' \alpha)$ measurement data from ref.~\cite{Kudeyarov}.
    If in the case of  lithium the $S_{L}(q)$ dependence plateau
    begins at $q = 2\: fm^{-1}$, the clusterization should be absente,
     as can be concluded from V.D. Efros calculation~\cite{Buki_1977}.}      }
    \item {Reasoning from the observed tendency of the $S_{L,max}$ decreasing
    with the growth of atomic number, in the case of lithium isotopes
    the $S_{L,max} \leq 0.9$ could be expected, but $S_{L,max} = 1.0$ was obtained.
    On the other hand the sum rule
    quenching ($S_{L,max} < 1.0$) can be explained by the nucleon modification
    inside the nuclear matter which have the density bigger than some critical value
    (see, for instance, ref.~\cite{Buki_2000}). Following this hypothesis,
    let us view the relation between $S_{L,max}$ and the nuclear matter density
    in the nucleus center ($\rho_0$). For $A \leq 3$ nuclei $S_{L,max}$ is equaled 1.0
    and $\rho_0 < 0.15$ nucleon/fm$^3$ and for the investigated $A \geq 4$ nuclei
     (besides $^{6,7}$Li nuclei) $S_{L,max}$ is less than 1.0 and $\rho_0 > 0.15$
     nucleon/fm$^3$. In case of $^{6,7}$Li $S_{L,max}$ is equal 1.0
     and $\rho_0 < 0.15$ nucleon/fm$^3$ similarly to $A \leq 3$ nuclei (though the atomic
    numbers of these nuclei are bigger than one of $^4$He).  }
\end{itemize}

Thus the obtained lithium isotope $S_{L}(q)$  values may be
considered as reason of the nucleon modification inside the nucleus
matter.
\end{multicols}

\vspace{5 mm}
\begin{multicols}{2}

\begin{center}

\end{center}
\end{multicols}
\end{document}